\documentclass[aip, reprint]{revtex4-1}

\usepackage{graphicx}
\usepackage{listings}
\usepackage[english]{babel}  
\usepackage{pifont}
\usepackage{amssymb}
\usepackage{amsfonts}
\usepackage{hyperref}
\usepackage[absolute]{textpos}

\begin{document}

\newcommand{\var}[1]{\emph{\ttfamily #1}}
\newcommand{\fun}[1]{{\bfseries\ttfamily #1}}
\newcommand{\syntaxItem}[1][] {\item[\ttfamily #1]}
\newenvironment{syntax}[0]{\begin{description}} {\end{description}}

\title{Conedy: a scientific tool to investigate Complex Network Dynamics} 
\author{Alexander \surname{Rothkegel}}
\email{alexander@rothkegel.de}
\affiliation{Department of Epileptology, University of Bonn, Sigmund-Freud-Str. 25, 53105 Bonn, Germany}
\affiliation{Helmholtz Institute for Radiation and Nuclear Physics, University of Bonn, Nussallee 14--16, 53115 Bonn, Germany}
\affiliation{Interdisciplinary Center for Complex Systems, University of Bonn, Br\"uhler Str. 164, 53175 Bonn, Germany}

\author{Klaus \surname{Lehnertz}}
\email{klaus.lehnertz@ukb.uni-bonn.de}
\affiliation{Department of Epileptology, University of Bonn, Sigmund-Freud-Str. 25, 53105 Bonn, Germany}
\affiliation{Helmholtz Institute for Radiation and Nuclear Physics, University of Bonn, Nussallee 14--16, 53115 Bonn, Germany}
\affiliation{Interdisciplinary Center for Complex Systems, University of Bonn, Br\"uhler Str. 164, 53175 Bonn, Germany} %
\date{\today}

\begin {abstract}
We present {\em Conedy}, a performant scientific tool to numerically investigate dynamics on complex networks. Conedy allows to create networks and provides automatic code generation and compilation to ensure performant treatment of arbitrary node dynamics. Conedy can be interfaced via an internal script interpreter or via a Python module.
\end {abstract}
\maketitle

\begin{textblock*}{20cm}(0.5cm,27cm)
The following article has been accepted by Chaos. After it is published, it will be found at \url{http://link.aip.org/link/?cha}.
\end{textblock*}

\begin{quotation}

Over the last decade, complex networks research has contributed significantly to improve our understanding of 
the dynamics of complex systems which are composed of mutually interacting components. The complicated relationship between network structure, dynamics of components, and emerging global behavior
requires tailored tools that aid in unraveling new phenomena and in verifying hypotheses and analytical results. Conedy (\url{http://www.conedy.org}) is a new scientific tool that allows to investigate dynamics on complex networks. In this paper, we give an overview of the functionality of Conedy and demonstrate its working procedure.
\end{quotation}

\section*{Introduction}
A network is generally considered as a collection of nodes, which are connected by edges. The generality and simplicity of this notion has established the description of complex systems with networks as a standard approach in many scientific disciplines \cite {Boccaletti2006a,BarratBook2008,Bullmore2009} that are concerned with a large number of functionally similar and mutually interacting components. Examples range from sociology and quantitative finance via biology, earth and climate sciences to physics and the neurosciences. The availability of computers allows for numerical investigations of such complex networks on far larger scales than by a manual approach. Both, structural and functional aspects of natural and man-made networks have been investigated, often considering thousands or millions of nodes. 

One approach to describe complex networks is to provide algorithms which generate model networks which are similar to complex networks in some aspect \cite {Watts1998, Barabasi1999, Newman2001b, Caldarelli2002, Catanzaro2005}. Apart from the merit of understanding structure and evolution of complex networks, such model networks have inspired many studies of dynamics on networks, in which each node is associated with a dynamical system. 
These studies include investigations of disease spreading \cite{Pastor-Satorras2001, ZhouTao2006, BarratBook2008, Stone2011}, of neural network dynamics \cite {Netoff2004, Percha2005, Dyhrfjeld-Johnsen2007, Feldt2007, Morgan2008, Rothkegel2009, Stratton2010}, of noise-induced phenomena \cite {Kwon2005, Mao-Sheng2006, Ren2010} and of synchronization phenomena \cite{WuBook2007,Arenas2008,SuykensOsipov2008,Lehnertz2009b,Almendral2011,Nishikawa2011}.

Edges in these networks typically indicate some kind of coupling, which is a weak disturbance of the node dynamics. The question arises in which way the global network dynamics can be related to both the node dynamics and to the structure of the network. Can we observe global signals, which are similar to those of single elements, or---even more interestingly---can we expect new emerging features which are not part of the dynamical spectrum of single elements? In order to untangle and understand the complicated relationship between network structure, global dynamics and node dynamics, a combined effort is necessary that includes theoretical approaches and network \cite{Albert2002,Newman2003,Costa2007} and time series analyses \cite{Abarbanel1993,Hegger1999,Kantz2003,Pikovsky_Book2001,Pereda2005,Hlavackova2007,Donner2010}. In this context, numerical studies play an important role in unraveling new phenomena, in verifying hypotheses and analytical results as well as in validating newly developed network and time series analysis techniques.

There are already software packages available that allow to integrate different types of dynamical systems, but they either focus on neuron dynamics \cite {Nusse1998, Ermentrout2002, Dhooge2003, Clewley2007} or they do not consider networks \cite {Bower1998, Neuron2006, Gewaltig2007, Goodman2009, Hines2009, Pecevski2009}. Here, we present Conedy, a computational tool which is aimed at scientists investigating dynamics on complex networks. Conedy allows to build arbitrary networks and provides generators for many classic and popular networks, such as lattices, random, small world, and scale-free networks. Conedy can handle dynamical systems which are ordinary differential equations, stochastic differential equations, iterated maps, and pulse-coupled oscillators. The dynamics of these systems can be assessed by a common interface, which easily allows to investigate the same network endowed with different dynamical systems as node dynamics.

For ease of use, Conedy can be interfaced with an inbuilt script interpreter or via Python bindings (\url{http://www.python.org}).  The inbuilt script interpreter may be useful for distributed computing. It provides support for creating Condor dagman job files, thereby converting the iterations of loops into different jobs. Condor  (\url{http://www.cs.wisc.edu/condor}) is a job management system developed at the Computer Science Department of the University of Wisconsin. Python is a free, open source, cross-platform programming language and has become a viable alternative to Matlab and Octave recently \cite{Oliphant2007}. The standalone Python bindings of Conedy put a powerful scientific tool at the fingertips of the analyst, which can then be combined  with an ever increasing number of other high-quality scientific packages for Python.  

Conedy is written in C++ and licensed under the GNU license (\url{http://www.gnu.org/copyleft/gpl.html}). Conedy is distributed as source code and as binary packages for Windows and Linux (\url{http://www.conedy.org}, \url{http://www.github.com/conedy/conedy}).  Conedy is not intended to contain plotting routines or support for other forms of visualization, which are excellently handled by other free software packages. 
Before we give an overview of the functionality of Conedy we provide a short example  to demonstrate its working procedure.

\section*{A Short Example}
\lstset{basicstyle=\ttfamily\footnotesize, frame=single}

In this example (written in Python), we simulate a small network of three coupled R\"ossler oscillators in the funnel regime, which display phase synchronization for a sufficiently large coupling strength \cite{Osipov2003,Nawrath2010}. The differential equations for an oscillator read:

\begin{eqnarray*}
\dot{x} &=& -  \omega y - z \\ 
\dot{y} &=& \omega x + a y  \\ 
\dot{z} &=& b + z (x - c) 
\end{eqnarray*}

\begin{figure}
\includegraphics[width=1.0\columnwidth]{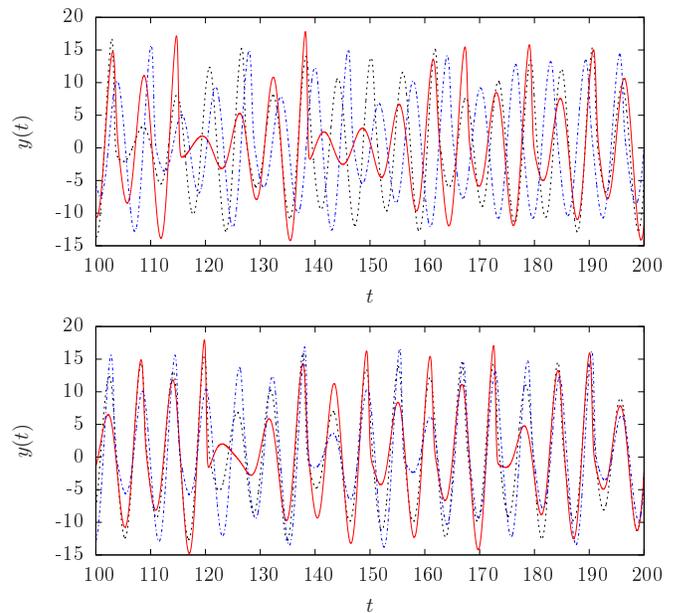}
\caption {(Color online) Second variable of three coupled R\"ossler oscillators in the funnel regime. Top: coupling strength $0.055$, bottom: coupling strength $0.2$.} \label{fig:shortexample}
\end{figure}

The oscillators are non-identical ($\omega_1=1.06$, $\omega_2=1.02$, $\omega_3=0.98$, $a=0.22$, $b = 0.1$, $c = 8.5$) and diffusively
coupled via their $y$-components. The dynamics is integrated with the default integrator and the second dynamical variable of the oscillators ($y$) is sampled in periodic intervals. We add three R\"ossler oscillators to an empty network
\begin{lstlisting}
import conedy as co
net = co.network()
co.set ("roessler_a", 0.22)
co.set ("roessler_b", 0.1)
co.set ("roessler_c", 8.5)
r1 = net.addNode(co.roessler())
r2 = net.addNode(co.roessler())
r3 = net.addNode(co.roessler())
net.setParam(r1, "roessler_omega", 1.06)
net.setParam(r2, "roessler_omega", 1.02)
net.setParam(r3, "roessler_omega", 0.98)
\end{lstlisting}
set random initial conditions,
\begin{lstlisting}
net.randomizeStates (co.roessler(), 
	co.uniform (-10.0, 10.0),
	co.uniform (-5.0, 5.0), 
	co.uniform (-0.5, 1.5))
\end{lstlisting}
connect oscillator 1 bidirectionally to both 2 and 3 with coupling strength 0.2,
\begin{lstlisting}
net.addEdge(r1, r2, co.weightedEdge(0.2))
net.addEdge(r2, r1, co.weightedEdge(0.2))
net.addEdge(r1, r3, co.weightedEdge(0.2))
net.addEdge(r3, r1, co.weightedEdge(0.2))
\end{lstlisting}
let 100 units of time pass to let transients die out, 
\begin{lstlisting}
net.evolve(0.0, 100.0)
\end{lstlisting}
select the time and each oscillator's second component (zero-based numbering) to be written to the file
\lstinline[basicstyle=\ttfamily]$roessler.dat$,
\begin{lstlisting}
net.observeTime("roessler.dat")
net.observe(r1, "roessler.dat", co.component(1))
net.observe(r2, "roessler.dat", co.component(1))
net.observe(r3, "roessler.dat", co.component(1))
\end{lstlisting}
set the sampling interval to 0.01, and observe the system for 100 units of time. 
\begin{lstlisting}
co.set("samplingTime", 0.01)
net.evolve(100.0, 200.0)
\end{lstlisting}
In Fig. \ref{fig:shortexample} we show the second variable of the oscillators as calculated with the listed script and, for comparison, for a coupling strength $0.055$. Phase synchronization can be observed for the
larger coupling strength. In the following, we will describe the design of Conedy and exemplify its working procedure.

\section*{Choosing Dynamics}
As most networks considered in both theoretical and experimental studies are sparse, Conedy administers an internal adjacency list representation for the investigated network, which consists of nodes and edges of different type. Each node type is associated with a dynamics, which can be of different kind. Conedy can handle dynamical systems of the following classes:
\begin{description}
\item[ordinary differential equations]
For numerical integration of ordinary differential equations, Conedy uses algorithms of the GNU Scientific Library (GSL) \cite{Galassi2009}, which contains step-controlled algorithms like the widely used Runge-Kutta-Fehlberg method of orders 4 and 5.
\item[stochastic differential equations]
Implementations of the Euler-Mayurama and the Milstein method are supplied \cite{KloedenPlaten1999}.
\item[pulse-coupled oscillators]
Integration is handled by an event-based approach \cite{Brette2007}. A relaxed heap  \cite {Driscoll1988} or a calendar queue \cite {Brown1988} can be chosen as priority queue at compile time. Interaction of oscillators  is defined by phase response curves \cite{Mirollo1990}. Conedy allows for delayed interactions and Poissonian excitations.
\item[iterated maps] Implemented straight-forwardly. 
\end{description}
Conedy ships with a list of some predefined node dynamics such as Kuramoto-, R\"ossler-, and Lorenz-oscillators or different neuron models \cite{Kuramoto1984,AlligoodSauerYorke1996,Izhikevich2007}. In case the desired dynamics is not in this list, Conedy supports automatic generation of code using a small description file, which follows the syntax of Windows INI files. In this description file, the dynamics and an integrator has to be specified. In addition, control parameters for the dynamics can be defined. After compilation of the code a new node type is available in Conedy. At run time, different values can be assigned to the control parameters for nodes of this type.

With the following example, we list and explain the description file, which would be required to add the dynamics of an excitable Barkley unit \cite{Barkley1991} to Conedy. The dynamics is often used in studies on pattern formation and is similar to that of a FitzHugh-Nagumo neuron.
The following stochastic differential equation describes the dynamics which is driven by Gaussian white noise \cite{perc2005}: 
\[
	dx_0 =  \frac{1}{k} x_0 ( 1 - x_0) ( x_0 - \frac {x_1 + b}{a}) dt  + \sigma dW, 
\]
\[
	dx_1 = (x_0 - x_1) dt.
\]
We start by defining a name for the node type:
\begin{lstlisting}
[gaussianBarkley]
\end{lstlisting}
We choose the type of the dynamical system as stochastic differential equations. The desired stepping algorithm can then be specified in the Python script.
\begin{lstlisting}
type = sde
\end{lstlisting}
Next, we specify the number of dynamical variables of the system,  here the number of differential equations:
\begin{lstlisting}
dimension = 2
\end{lstlisting}
We specify the number of control parameters for the dynamics: 
\begin{lstlisting}
parameter = 4
\end{lstlisting}
Each control parameter is assigned a name and a default value.
\begin{lstlisting}
parametername1 = a
defaultvalue1 = 0.75
parametername2 = b
defaultvalue2 = 0.01
parametername3 = k
defaultvalue3 = 0.05
parametername4 = sigma
defaultvalue4 = 0.18
\end{lstlisting}
Finally, we define the dynamics of the node type, i.e., the derivatives $\frac{dx_0}{dt}$ and $\frac{dx_1}{dt}$ as well as the noise terms $s_0 = \frac{dx_0}{dW}$ and $s_1 = \frac{dx_1}{dW}$.  
\begin{lstlisting}
dynamics=
  dxdt[0] = 1/k*x[0]*(1-x[0])*(x[0]-(x[1]+b)/a);
  dxdt[1] = x[0] - x[1];
  s[0] = sigma;
  s[1] = 0.0;
  dsdx[0] = 0.0;
  dsdx[1] = 0.0;
\end{lstlisting}
As the noise is not multiplicative, the derivatives $\frac{ds_0}{dx_0}$ and $\frac{ds_1}{dx_1}$ can be omitted. 
The equations use C-syntax and may contain---in addition to standard C-constructs---mathematic functions as defined in \lstinline[basicstyle=\ttfamily]$math.h$.

While nodes are associated with dynamics, edges indicate some kind of coupling. The exact nature of this coupling depends on the node that the edge connects to. For example, for an iterated map the edge may signify the typical coupling for a coupled map lattice, while it may indicate diffusive coupling for ordinary differential equations. 
The coupling has to be defined alongside the internal dynamics of the node type. For this purpose the node is provided (by each edge) with two numbers, which are interpreted as \lstinline[basicstyle=\ttfamily]$state$ of the coupled node and as coupling \lstinline[basicstyle=\ttfamily]$weight$. Edges of different type differ in the way they determine these two numbers. The weight may be stored per edge or only per edge type to save memory.  The state may be---depending on the edge type---a variable of the coupled node (e.g. $x_0$) or some transformation of a variable. 

To define the coupling, the macro \lstinline[basicstyle=\ttfamily]$forEachEdge$ can be used, which loops over all adjacent edges, sets coupling weight and state to the values provided by this edge, and performs an instruction which is given as argument to the macro. To include diffusive coupling to the definition of the Barkley dynamics, we add the following line to the property \lstinline[basicstyle=\ttfamily]$dynamics$ of the description file:

\begin{lstlisting}
forEachEdge
  (dxdt[0] = dxdt[0] + weight*(state-x[0]);)
\end{lstlisting}
The new node type is added to Conedy by depositing the description file into a directory which is monitored by Conedy. For sake of performance, Conedy has to be recompiled after each addition of a new node type. Under Linux this is handled automatically at the next import of Conedy to the Python interpreter. 

\section*{Building Networks}
In the following we explain the different approaches for network generation and manipulation available in Conedy.
For the sake of simplicity, we assume that Conedy has been imported to Python in the standard way by 
\begin{lstlisting}
import conedy as co
\end{lstlisting}
and that a network has been declared by
\begin{lstlisting}
net = co.network()
\end{lstlisting}
At this point, the network \lstinline[basicstyle=\ttfamily]$net$ is empty and  serves as a starting point for the various manipulation functions which allow to build the desired network step by step.

Most of the functions in Conedy which manipulate networks accept node templates and edge templates as arguments. These are Python classes, which specify node/edge type and values of control parameters. 
To create a node template of a certain type using default control parameters, Conedy accepts the following syntax:
\begin{syntax}
\syntaxItem [\var{nodeType}()]
Return a template of node type \var{nodeType} using default values for control parameters.
\end{syntax}
Alternatively, it is possible to create a node template with user-defined values for control parameters: 
\begin{syntax}
\syntaxItem[\var{nodeType}(\var{p1}, ..., \var{pM})]
Return a template of type \var{nodeType} using \var{p1}, ..., \var{pM} as values for the $M$ control parameters the node type depends on.
\end{syntax}
Templates for edges can be created in a similar way. 
Conedy supplies elementary functions, which add a single node or a single edge to the network:
\begin{syntax}
\syntaxItem[\var{net}.\fun{addNode}(\var{nt})]   
Add a node to the network \var{net} according to the node template \var{nt} and return a unique identifying integer  (the node number \var{n}) for the added node.
\syntaxItem[\var{net}.\fun{addEdge}(\var{s}, \var{t}, \var{et})]
Add an (directed) edge to the network \var{net} according to the edge template \var{et} connecting node \var{s} with node \var{t}.
\end{syntax}
With the help of these elementary functions it is possible to create arbitrary networks in Python.  For building commonly used networks such as random networks or lattices, higher level functions are available, e.g.: 
\begin{syntax}
\syntaxItem[\var{net}.\fun{randomNetwork}(\var{N}, \var{p}, \var{nt}, \var{et})]
	Add \var{N} nodes according to the node template \var{nt} to the network \var{net}.  Connect each pair of newly added nodes with probability \var{p} by an edge according to the  edge template \var{et}. Return the node number \var{n} of the first added node. Other added nodes have consecutive numbers starting with \var{n}+1.
\end {syntax}
\begin{syntax}
\syntaxItem[\var{net}.line(\var{N}, \var{k}, \var{nt}, \var{et})]
Add \var{N} nodes according to the node template \var{nt} to the network \var{net}. Connect the newly added nodes in order to form an open chain in which nodes (except the ones near the boundaries) are connected to their \var{k} nearest neighbors to each side. Return the node number \var{n} of the first added node. Other added nodes have consecutive numbers starting with \var{n}+1.
\end {syntax}

In addition, small-world or random networks with a predefined degree distribution can be created using manipulation functions that allow to add (or to replace existing edges by) random edges. 
Finally, it is possible to create networks from adjacency lists or matrices stored in external files. To verify and investigate the created network, some standard network measures from network theory like clustering coefficient, average shortest path length and centralities are implemented. In addition, Conedy provides export functions to transfer networks to other network analysis tools such as igraph \cite{Csardi2006} or NetworkX \cite{Hagberg2008}.

\section*{Setting control parameters and initial conditions}
For networks which consist of nodes with one or a few sets of control parameters and initial conditions, it is convenient to set these when defining the node templates. 
Additionally, it is possible to draw control parameters and initial conditions from random distributions (uniform, Gaussian,  exponential). Random number generation for this (and for the stochastic integration) is handled by the GSL. Both algorithm and random seed can be controlled to allow for reproducible computations. 
\begin{syntax}
\syntaxItem [ \var{net}.\fun{randomizeParameter}(\var{paramName}, \var{ranDist}) ]
Draws for every node---which is of the node type to which control parameter \var{paramName} belongs---a random value for \var{paramName} from the distribution \var{ranDist}.
\end{syntax}
In a similar way, initial conditions of nodes in the network can be drawn from random distributions.
\begin{syntax}
\syntaxItem[ \var{net}.\fun{randomizeStates}(\var{nt}, \var{ranDist}\{, \var{ranDist} \})]. Draw a random number for every dynamical variable of all nodes which match the node template \var{nt}. 
For every dynamical variable, a distribution \var{ranDist} has to be chosen.
\end{syntax}

\section*{Choosing Observables and Starting Numerical Integration}
Conedy writes the values of one or more selected observables of the network (e.g., first dynamical variable of the second node, third dynamical variable of the first node, etc.) to one or several files. In addition, observables such as mean network activity or Kuramoto's order parameter \cite{Kuramoto1984} can be selected. Conedy administers a list of observables for every file and orders the columns in these files in the same order as \fun{observe} commands have been issued:
\begin{syntax}
\syntaxItem [\var{net}.\fun{observe} (\var{n}, \var{fileName}, \var{et}) ]
	Add the state of node \var{n}---as returned by an edge according to template \var{et}---to the list of observables to be written to file \var{fileName}.
\end{syntax}
The standard output is a matrix with rows corresponding to sampling times and columns corresponding to observables of the network.  Eventually, a call of the \fun{evolve} function starts the integration, thereby sampling all registered observables with a globally defined sampling interval:
\begin{syntax}
\syntaxItem [\var{net}.\fun{evolve} (\var{s}, \var{t})]
	Set the time to \var{s} and evolve the dynamics on the network \var{net} until the time \var{t}.
\end{syntax}
Data can be written to whitespace-newline-separated text files, bzip-compressed text files, or to binary files.  

\section*{Example \#1: Small-world network of diffusively coupled stochastic Barkley units}

\begin{figure}
\includegraphics[width=\columnwidth]{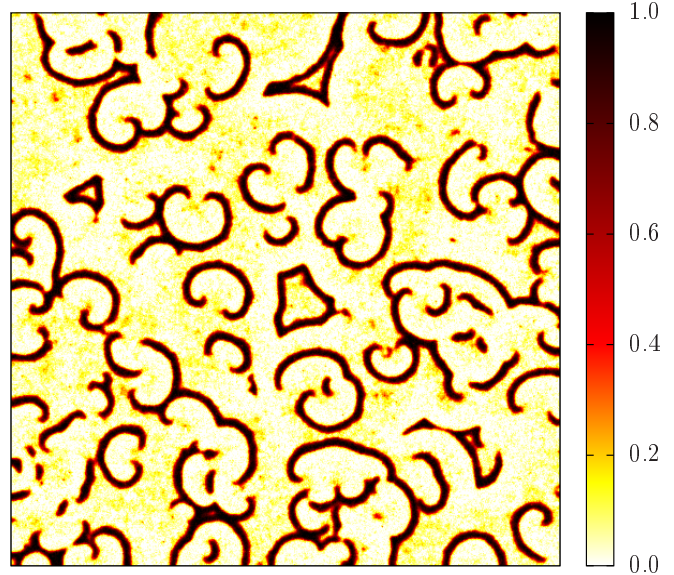}
\caption{(Color online) Spiral waves in a small-world network of $512 \times 512$ diffusively coupled noisy Barkley units (values range from $0.0$ (white) to $1.0$ (black)). Snapshot of the spatial distribution of the first dynamical variable.\label{spiralWave}  }
\end{figure}

As a first usage example, we consider the formation of wave patterns in a two-dimensional small-world network of diffusively coupled Barkley units driven by Gaussian white noise~\cite{Barkley1991,perc2005}. For this purpose we use the node type defined above. We begin with creating a node template and specify its initial conditions as the fixed point at ($x_0=y_0=0.0$).
\begin{lstlisting}
bark = co.gaussianBarkley()
bark.setState (0.0, 0.0)
\end{lstlisting}
Next, we create a two-dimensional lattice of $512 \times 512$ units, where each pair of units is connected if their Euclidean distance is smaller than or equal to 1.0 (nearest neighbor coupling). The coupling strength for each edges is 3.84. The network is
then rewired by replacing one per mill of edges with edges between randomly chosen source and target nodes.
\begin{lstlisting}
net.lattice(512, 512, 1.0, bark, co.weightedEdge (3.84))
net.rewire(0.001)
\end{lstlisting}
By applying a Milstein method to the resulting 524288-dimensional stochastic differential equation system, we let the time evolve to 20.0. This is sufficient to let transients die out.
\begin{lstlisting}
net.evolve(0.0, 20.0)
\end{lstlisting}
Next, we add the first dynamical variable of all nodes to the list of observables and specify the name of the file as \lstinline[basicstyle=\ttfamily]$waves.dat$.
\begin{lstlisting}
net.observeAll("waves.dat", co.component(0))
\end{lstlisting}
Finally, we generate a snapshot of all registered observables at the current time (20.0) of the network: 
\begin{lstlisting}
net.snapshot()
\end{lstlisting}
The data is then written to the file \lstinline[basicstyle=\ttfamily]$waves.dat$ and is shown in Fig. \ref{spiralWave}.
At the time of the snapshot, spiral waves can be observed from the spatial distribution of the first dynamical variable of the units.

\section*{Example \#2: Random network of pulse-coupled oscillators}
As a second usage example we consider chaotic transients to complete synchrony in a random network of pulse-coupled oscillators (PCOs)~\cite{Zumdieck2004,Rothkegel2011}. PCOs are described by their phase $\phi \in [0,1]$ with $\dot{\phi} = 1$. When $\phi$ reaches $1$ the oscillator fires and $\phi$ is reset to 0.
The interaction between PCOs is defined by the phase response function $\Delta (\phi)$, i.e., the dependence of the phase change $\Delta$ on the phase $\phi$ of an oscillator due to the firing of a connected oscillator.  

\begin{figure}
\includegraphics[width=\columnwidth]{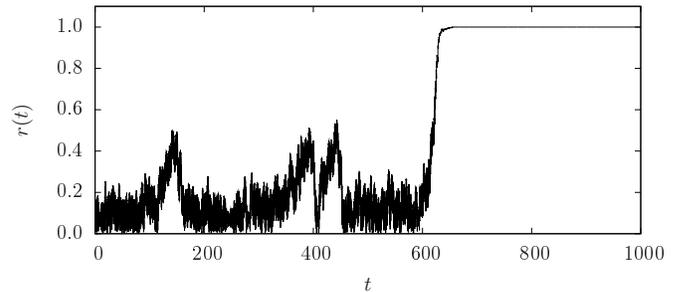}
\caption{Spontaneous transition to complete synchrony in a random network of 1000 pulse-coupled oscillators. Temporal evolution of Kuramoto's order parameter $r(t)$.\label{order} }
\end{figure}

Whenever at some time~$t$ an oscillator receives a pulse, its phase is updated according to
\[
	\phi(t^+) = \phi(t) + \Delta (\phi).
\]
Here, we consider the case of a phase response function $\Delta(\phi)$ which is limited to the value that is needed for the excited oscillator to fire immediately~\cite{Mirollo1990}. \[
	\Delta (\phi) = \min \{a + b \phi, 1 - \phi \},  \phi \in (0,1).
\]
We assume that oscillators are not excitable at the time of firing ($\Delta (0) = \Delta (1) = 0)$. The dynamics of these PCOs can be integrated into Conedy by the following description file. As above we declare name, integrator, dimension, and control parameters of a node:
\begin{lstlisting}
[pcoMirollo]
type = pco
dimension = 1
parameter = 2
parametername1 = a
defaultvalue1 = 0.01
parametername2 = b
defaultvalue2 = 0.02
\end{lstlisting}
Instead of a differential equation, the property \lstinline[basicstyle=\ttfamily]$dynamics$ now contains the phase change \lstinline[basicstyle=\ttfamily]$delta$ in dependence on the phase (\lstinline[basicstyle=\ttfamily]$phase$):
\begin{lstlisting}
dynamics=
  if (phase == 0) {
  	delta = 0;
  }
  else {
  	delta = a + b*phase;
  	if (delta + phase > 1.0)
  		delta = 1.0 - phase;
  }
\end{lstlisting}
Next, we create a random network of PCOs with $N=1000$ nodes of type \lstinline[basicstyle=\ttfamily]$pcoMirollo$, where nodes are connected with a probability of 0.01 by unweighted edges.
\begin{lstlisting}
net.randomNetwork(1000, 0.01, 
	co.pcoMirollo(),
	co.edge())
\end{lstlisting}
Initial conditions are drawn from the uniform distribution over the interval $[0,1]$:
\begin{lstlisting}
net.randomizeStates(co.pcoMirollo(), 
	co.uniform(0.0, 1.0))
\end{lstlisting}
We reset the default control parameters for the oscillators:
\begin{lstlisting}
co.set ("pcoMirollo_a", 0.015)
co.set ("pcoMirollo_a", 0.045)
\end{lstlisting}
As observables we register the sampling time and Kuramoto's order parameter $r(t) = 1/N \left|\sum_{n} e^{2 \pi i \phi_n(t)}\right|$ for all $N$ nodes. $r(t)$ yields 1 for complete synchrony and 0 for asynchronous firing. Both observables are to be written to the file \lstinline[basicstyle=\ttfamily]$order.dat$, and we let the time evolve 1000.0 time units.
\begin{lstlisting}
net.observeTime("order.dat")
net.observePhaseCoherence("order.dat")
net.evolve(0.0, 1000.0)
\end{lstlisting}
Fig. \ref{order} shows the temporal evolution of the order parameter $r(t)$, which exhibits a spontaneous transition to synchrony. 

\section*{Performance considerations} 
In scientific computing, it is often possible to find specialized solutions, which perform well for certain conditions but either have a narrowed functionality or perform worse for other conditions. 
For numerical integration of complex network dynamics, different kinds of optimization can be achieved if the network is required to be in some way homogeneous. A more general approach will allow for different edge/node types and control parameters. 

\subsection*{Homogeneity of node and edge types}
A simulation tool for dynamics on complex networks that consist of different edge/node types (mixed network) can be realized by making use of virtual functions and jump tables to determine the edge types at run-time. This also offers the convenient possibility to change the type of edges without recompilation of the source code. At the downside, however, the use of virtual functions may be unnecessary slow for networks consisting of identical types for all edges of a node (homogeneous networks). In addition, the jump tables for virtual functions consume a considerable amount of memory. 

To enable an efficient handling of both situations (homogeneous and mixed networks), Conedy follows two different approaches to the internal representation of edges. First, Conedy allows for nodes that can connect to edges of arbitrary type (mixed networks) by making use of virtual functions. The second approach enables an efficient handling of homogeneous networks. For this purpose, Conedy allows for nodes, for which the connecting edge type is specified at compile time. Numerical integration can be achieved without virtual function calls using these nodes. To create a specialized node type, which connects to edges of a predefined type, we specify this edge type in the description file. For example, we complement the description file for the Barkley dynamics by:
\begin{lstlisting}
staticEdgeType=component_weightedEdge
\end{lstlisting}
With this example, nodes of type \lstinline[basicstyle=\ttfamily]$gaussianBarkley$ are connected by edges of type \lstinline[basicstyle=\ttfamily]$component_weightedEdge$ only.

\begin {figure}
\includegraphics[width=\columnwidth]{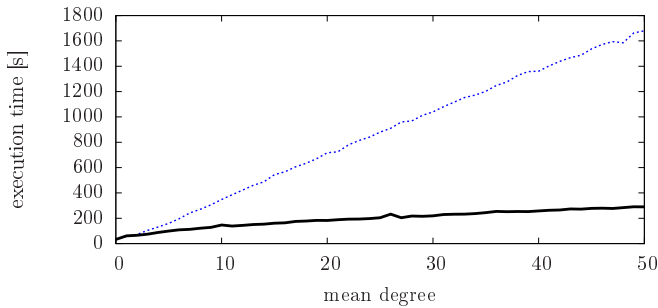}
\caption{(Color online) Execution times in dependence on the mean degree of random networks of 5000 R\"ossler oscillators with eigenfrequency 0.89. Values were obtained using nodes allowing for inhomogeneous node/edge types (blue dotted line) and using nodes which force connecting edge types at compile time (black solid line) on a PC with 1.5 GFLOPS (2.4 GHz). Integration method: Runge-Kutta-Fehlberg with step size control and an absolute error of $10^{-5}$.
The dynamics of the networks was integrated for 200.0 time units. (Lines are for eye-guidance only). \label{compareStaticVirtual} }
\end{figure}

With the second approach the memory consumption of unweighted edges in homogeneous networks can be reduced by a factor of four \footnote {Conedy uses a 32 bit integer as identifier for the target node of the edge. For edges with virtual functions, 64 bit is needed for the jump table and most modern compiler will leave another 32 bit empty due to alignment of the data structures.}. To compare both approaches, we numerically integrate the dynamics of a random network of 5000 R\"ossler oscillators. In Fig.~\ref{compareStaticVirtual} we show the respective execution times in dependence on the mean degree: the higher the mean degree, the greater the performance benefit of the approach specialized for homogeneous networks.

\subsection*{Homogeneous node parameters}
For networks consisting of a large number of identical nodes with identical control parameters, it is not desirable to reserve memory for the parameters of each node.
However, a specialized implementation for fixed parameters will narrow functionality. To provide a reasonable implementation for both homogeneous and inhomogeneous node parameters, Conedy uses parameter sheets, which are stored in a lookup table. Each node reserves a 32 bit integer for the number of the sheet which contains the node's parameters. Nodes with identical parameters share the same sheet number which reduces memory consumption.
When parameters of nodes are drawn from a random distribution, each node is assigned its own parameter sheet. Note, that in this situation, the chosen implementation will lead to a slightly increased memory consumption as compared to a direct storage of control parameters as part of the nodes.

\subsection*{Event-driven integration of pulse-coupled oscillators}
Conedy handles the numerical integration of pulse-coupled oscillators (PCOs) via an event-driven approach \cite{Brette2007}, which makes use of a priority queue. This queue contains the PCOs ordered by the time of their next firing. When an oscillators fires, it is  removed from the top of the queue and reinserted at the bottom. In addition, the priority (firing time) of all connected oscillators is decreased/increased. Depending on the implementation of the queue, these four operations (remove, insert, increase priority, decrease priority) may scale differently with the size of the queue. In principle, with the use of a calendar queue \cite {Brown1988}, constant scaling for all four operations is possible. However, this queue has a comparably large overhead and only performs well, if events are distributed homogeneously in time, which corresponds to asynchronous oscillators. Alternatively, heap structures can be used which are equally well suited for arbitrary event distributions. They lack, however, an efficient way to decrease the priority of events (which is necessary to handle negative phase responses).

\begin{figure}
\includegraphics[width=\columnwidth]{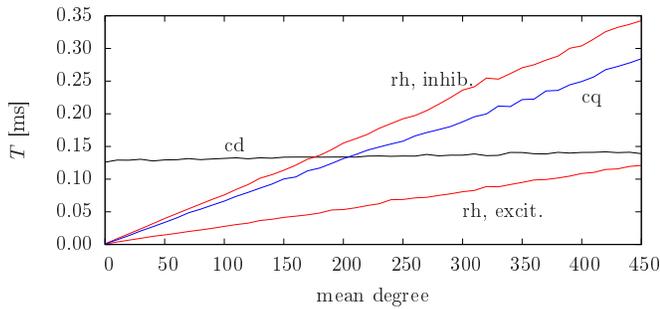}
\caption{
(Color online) Execution times per firing neuron $T$ in dependence on the mean degree for random networks of 10000 integrate-and-fire neurons.
Values were obtained using event-driven approaches implemented in Conedy with a calendar queue (``cq'', blue), with a relaxed heap (``rh'', red) and with a clock-driven approach (``cd'', black; Runge-Kutta-Fehlberg with step size control and an absolute error of $10^{-5}$). The calculation was done for both inhibitory and excitatory neurons. While the relaxed heap implementation is much faster for excitatory neurons (``rh, excit.''), both types of neurons are integrated equally fast with the calendar queue and the clock-driven integrator. Tests were performed on a PC with 1.5 GFLOPS (2.4 GHz). (Lines are for eye-guidance only).} 
\label{compareEventDgl} 
\end {figure}

In Conedy, both a relaxed heap and a calendar queue can be chosen at compile time. Compared to a clock-driven approach (e.g., a Runge-Kutta scheme), the integration is exact (as far as admitted by double precision) and does not require the choice of a step size.   To compare an event-driven with a clock-driven integrator we 
measure execution times for random networks of either 10000 excitatory or 10000 inhibitory pulse-coupled integrate-and-fire neurons \cite {Olmi2010, Jahnke2009}. In Fig.~\ref {compareEventDgl} we show execution times in dependence on the mean degree \footnote{We assume that neurons are described by an membrane potential $v(t)$, which is governed by $dv(t)/dt 
= - a v(t) + b$, while every presynaptic firing neuron at time $t_f$ induces a jump $ v(t_f^+) = v(t_f) \pm c$}. We chose control parameters, such that neurons fire asynchronously. The relaxed heap outperforms the calendar queue for excitatory oscillators, while the calendar queue seems better suited for inhibitory oscillators. For large mean degrees, the event-driven integration is in both cases outperformed by a clock-driven approach, in which we check for firing oscillators after every integration step. This is because in the phase description, excitations of integrate-and-fire neurons are computationally more costly as they require the evaluation of exponentials.   

\section*{Conclusion}
Conedy is a convenient, easy-to-use scientific tool for people interested in investigating dynamics on complex networks. It allows to create networks via elementary functions (which add single nodes and edges), via higher level functions for often investigated networks (e.g., lattices, random, small world, and scale-free networks), or from external data (in the form of adjacency matrices or lists). The size of a network is limited by the available computer memory only. Conedy provides automatic code generation and compilation to ensure performant treatment of arbitrary node dynamics. We here have concentrated on the Python interface of Conedy and on usage examples.

As algorithms for numerical integration have become complex and at the same time very reliable, it is prudent to hide much of their complexity behind frameworks to clear the sight for the system under study. Conedy combines several integration schemes that are summarized by a common interface, which easily allows to investigate the same network endowed with different kinds of dynamical systems as node dynamics. As with similar scientific software packages \cite{Bower1998, Nusse1998, Ermentrout2002, Dhooge2003, Neuron2006, Clewley2007, Gewaltig2007, Goodman2009, Hines2009, Pecevski2009}, it is advisable that the potential user has a thorough background in the numerical treatment of dynamical systems which cannot be provided by this paper, but rather by textbooks \cite{KloedenPlaten1999,Press2007}. The user will still need considerable experience in order to avoid spurious interpretation of the results, especially for large networks.

The last years have seen an extraordinary success of complex network and their applications in diverse disciplines. We hope that with Conedy we can contribute to this rapidly evolving field.

\section*{Acknowledgments}
We are grateful to Gerrit Ansmann, Stephan Bialonski, Henning Dickten, Justus Schwabedal, Ferdinand Stolz, and Tobias Wagner for their contributions to the source code and documentation of Conedy and for valuable comments on earlier versions of the manuscript. This work was supported by the Deutsche Forschungsgemeinschaft (LE 660/4-2).

\end{document}